\documentclass[a4paper]{jpconf}
\usepackage{graphicx}
\newcommand{\qq}{Q'} 
\newcommand{\re}{\mathrm{Re}\,}

\newcommand{\gev}{\,{\rm GeV}}

\begin{document}
\title{Exploring the nucleon structure through GPDs and TDAs in hard exclusive processes}

\author{ B.~Pire$^1$, K.~Semenov-Tian-Shansky$^{1,2}$, 
 L.~Szymanowski$^{3}$ and J. Wagner$^3$}

\address{$^1$ CPhT, \'Ecole Polytechnique,
CNRS, 91128 Palaiseau,     France}
\address{$^2$ LPT, Universit\'e Paris-Sud, CNRS, 91405 Orsay, France}
\address{$^3$ So{\l}tan Institute for Nuclear Studies,
Ho\.za 69, 00-681 Warsaw, Poland}

\ead{pire@cpht.polytechnique.fr}

\begin{abstract}
Generalized Parton Distributions (GPDs) offer a new way to access the quark and gluon nucleon structure. We review recent progress in this domain, emphasizing the need to supplement the experimental study of deeply virtual Compton scattering by its crossed version, timelike Compton scattering. We also describe the extension of the GPD concept to three quark operators and the relevance of their nucleon to meson matrix elements, namely the transition distribution amplitudes (TDAs) which factorize in backward meson electroproduction and related processes. We discuss the main properties of the TDAs.
\end{abstract}

\section{Introduction}
The study of the internal structure of the nucleon has been the subject of many developments in the past decades and the concept of generalized parton distributions has allowed a breakthrough in the 3 dimensional description  of the quark and gluon content of hadrons. Hard exclusive reactions have been demonstrated to allow to probe the 
quark and gluon content of protons and heavier nuclei.

In this short review, we concentrate on two timely items: firstly, we emphasize the complementarity of timelike and spacelike studies of hard exclusive processes, taking as an example the case of timelike Compton scattering (TCS) where data at medium energy should be available at JLab@12 GeV, supplemented by higher energy data thanks both to a forthcoming electron-ion collider and to the study of ultraperipheral collisions at the LHC. Secondly, we show how the backward region of deeply virtual meson electroproduction processes may be described in a factorized way, with  the necessary extension of the GPD concept to three quark operators and  their nucleon to meson matrix elements,  the transition distribution amplitudes (TDAs). 

\section{Timelike vs spacelike DVCS}
A considerable amount of theoretical and experimental work has 
been devoted to the study of deeply virtual Compton scattering (DVCS),
 i.e., $\gamma^* p \to \gamma p$, 
an exclusive reaction where generalized parton
distributions (GPDs) factorize from perturbatively calculable coefficient functions, when
the virtuality of the incoming photon is large enough~\cite{historyofDVCS}.
It is now recognized that the measurement of GPDs should contribute in a decisive way to
our understanding of how quarks and gluons assemble into
hadrons~\cite{gpdrev}. In particular, the transverse
location of quarks and gluons become experimentally measurable via the transverse momentum dependence of  GPDs \cite{Burk}. Results on DVCS \cite{DVCSexp} obtained at HERA and JLab already allow to get a rough 
idea of some of the GPDs (more precisely of the Compton form factors) in a restricted but 
interesting kinematical domain \cite{fitting}. An extended research program at JLab@12 GeV and 
Compass is now proposed to go beyond this first set of analysis. 
This will involve taking into account  next to leading order in $\alpha_s$ and next to 
leading twist contributions \cite{APT}.
 
 The physical process where to observe the inverse reaction,  timelike Compton scattering (TCS) \cite{BDP},
 \begin{equation}
 \gamma(q) N(p) \to \gamma^*(q') N(p')
 \end{equation}
 is   the exclusive photoproduction of a
heavy lepton pair, $\gamma N \to \mu^+\!\mu^-\, N$ or $\gamma N \to
e^+\!e^-\, N$, at small $t = (p'-p)^2$ and large \emph{timelike} final state lepton pair squared mass $q'^2 = Q'^2$; TCS 
shares many features with DVCS. The generalized Bjorken variable in that case is $\tau = Q'^2/s $
 with $s=(p+q)^2$. One also defines $\Delta = p' -p$  ($t= \Delta^2$) and the skewness variable 
  $\eta$ as
$$\eta = - \frac{(q-q')\cdot (q+q')}{(p+p')\cdot (q+q')} \,\approx\,
           \frac{ Q'^2}{2s  - Q'^2} = \frac{ \tau}{ (2-\tau)}.$$
At the Born order, the TCS amplitude is described by the handbag diagrams of Fig. 1.
As in the case of DVCS,  purely electromagnetic
mechanism where the lepton pair is produced through the Bethe-Heitler (BH) subprocess 
$\gamma (q)\gamma^* (\Delta) \to\ell^+\ell^-\;,$  contributes at the amplitude level. This amplitude 
is completely calculable in QED provided one knows the  nucleon form factors at small $\Delta^2 = t$.
This process has a very peculiar angular dependence and overdominates the TCS process if
one blindly integrates over the final phase space. One may however choose kinematics where 
the amplitudes of the two processes are of the same order of magnitude, and either subtract the 
well-known Bethe-Heitler process or use specific observables sensitive to
 the interference of the two amplitudes. Finally some kinematical cuts may allow to decrease 
 sufficiently the Bethe Heitler contribution.

\begin{figure*}
\includegraphics[width=0.4\textwidth]{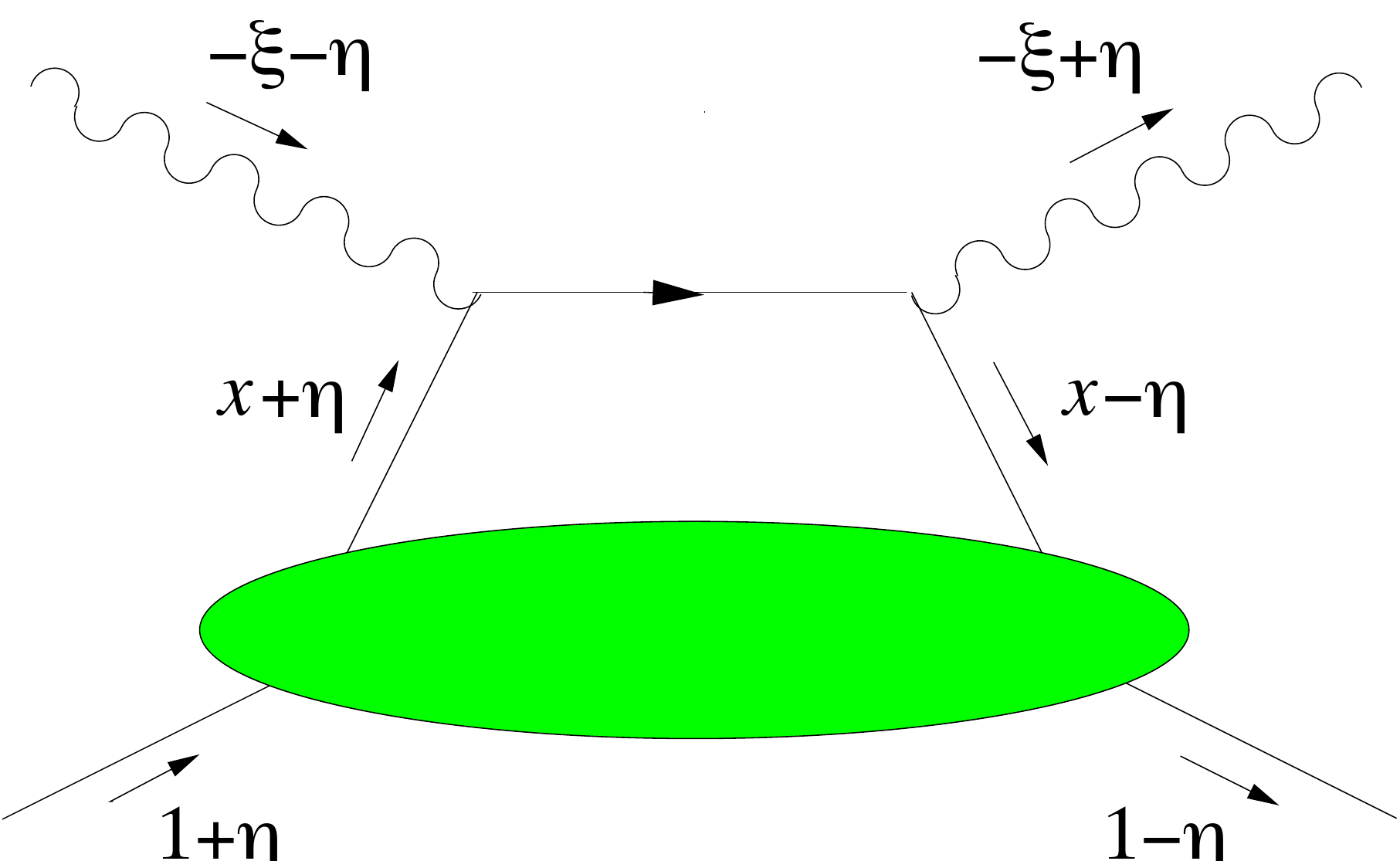}
\hspace*{2cm} \includegraphics[width=0.4\textwidth]{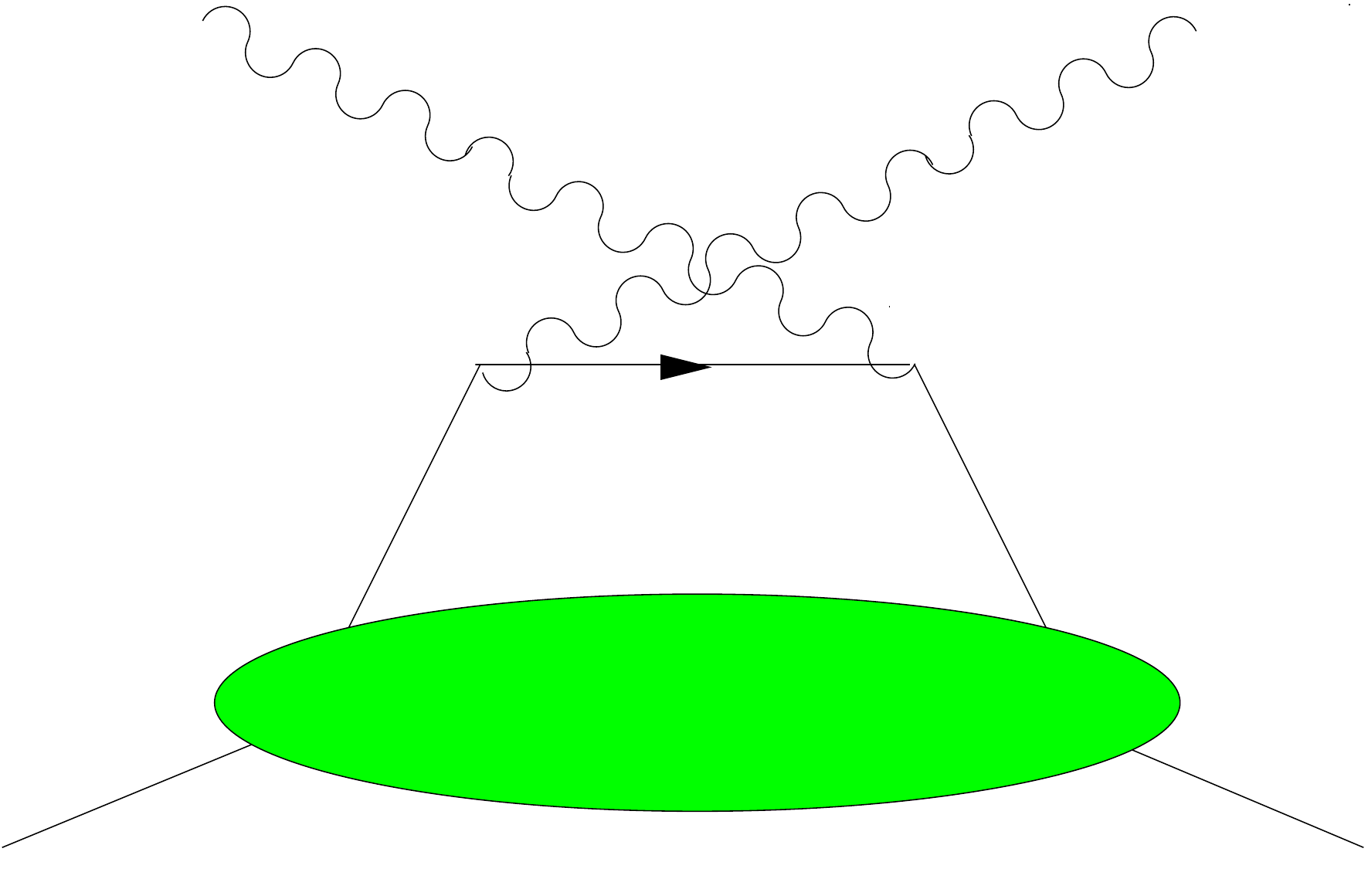} 
\caption{Handbag diagrams for the Compton process in the scaling limit. The
plus-momentum fractions $x$, $\xi$, $\eta$ refer to the average proton
momentum $\frac{1}{2}(p+p')$. In the DVCS case, $\xi = \eta$ while in the TCS case $\xi = -\eta$.}
\label{fig:1}       
\end{figure*}
 
\begin{figure}
\hspace*{2cm} \includegraphics[width=12cm]{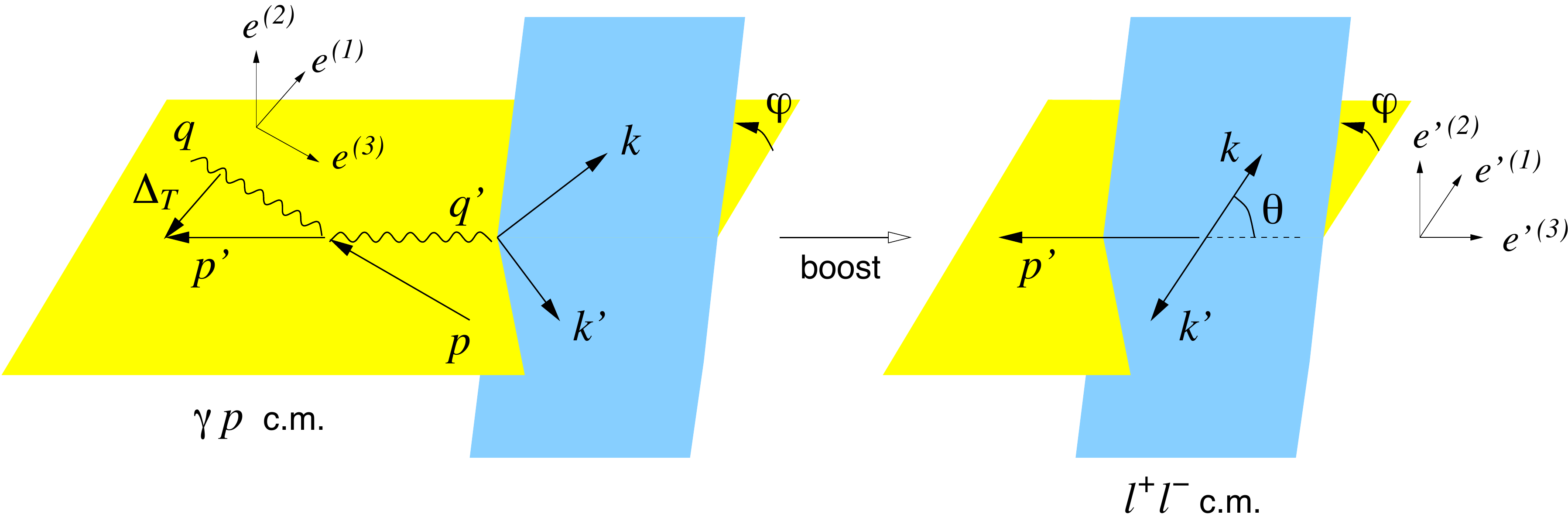}
\caption{Kinematical variables and coordinate axes in
the $\gamma p$ and $\ell^+\ell^-$ c.m.\ frames.}
\label{angle}      
\end{figure}

The kinematics of the $\gamma (q) N (p)\to \ell^-(k) \ell^+(k') N(p')$ process is shown in Fig.\ref{angle}.
 In the $\ell^+\ell^-$ center of mass system,  one introduces the polar and azimuthal angles $\theta$
and $\varphi$ of $\vec{k}$, with reference to a coordinate system with
$3$-axis along $-\vec{p}\,'$ and $1$- and $2$-axes such that $\vec{p}$
lies in the $1$-$3$ plane and has a positive $1$-component.
Since the amplitudes for the Compton and Bethe-Heitler
processes transform with opposite signs under reversal of the lepton
charge,  the interference term between TCS and BH is
odd under exchange of the $\ell^+$ and $\ell^-$ momenta.
It is thus possible to
 project out
the interference term through a clever use of
 the angular distribution of the lepton pair. 
The interference part of the cross-section for $\gamma p\to \ell^+\ell^-\, p$ with 
unpolarized protons and photons has a characteristic ($\theta, \varphi$) dependence given  by (see details in \cite{PSW})
\begin{eqnarray}
   \label{intres}
\frac{d \sigma_{INT}}{d\qq^2\, dt\, d\cos\theta\, d\varphi}
= {}-
\frac{\alpha^3_{em}}{4\pi s^2}\, \frac{1}{-t}\, \frac{M}{Q'}\,
\frac{1}{\tau \sqrt{1-\tau}}\,
  \cos\varphi \frac{1+\cos^2\theta}{\sin\theta}
     \re{\cal M} \; ,\nonumber
\end{eqnarray}
with 
\begin{equation}
\label{mmimi}
{\cal M} = \frac{2\sqrt{t_0-t}}{M}\, \frac{1-\eta}{1+\eta}\,
\left[ F_1 {\cal H}_1 - \eta (F_1+F_2)\, \tilde{\cal H}_1 -
\frac{t}{4M^2} \, F_2\, {\cal E}_1 \,\right],
\nonumber
\end{equation}
where $-t_0 = 4\eta^2 M^2 /(1-\eta^2)$, ${\cal H},  \tilde{\cal H}, {\cal E}$ are Compton form factors and $F_1, F_2$ are the nucleon Dirac and Pauli form factors.
 With the integration limits symmetric about $\theta=\pi/2$ the interference
term changes sign under $\varphi\to \pi+\varphi$ due to charge conjugation,
whereas the TCS and BH cross sections do not. One may thus extract the 
Compton amplitude through a study of
$\int\limits_0^{2\pi}d\phi\,\cos \phi \frac{d\sigma}{d\phi}$.

 This program has not yet been experimentally successful \cite{NadelTuronski:2009zz} due to the existing limited quasi real photon flux in the appropriate kinematical domain both at JLab and HERA. This will be  much improved with the JLab@12 GeV program, both in Hall B \cite{Albayrak} and in Hall D. These experiments will enable to test the universality of GPDs extracted from DVCS and from TCS, provided NLO corrections are taken into account. Experiments at higher energies, e.g. in ultraperipheral collisions at RHIC and LHC \cite{PSW} or in a future electron-ion collider \cite{eic}, may even become sensitive to gluon GPDs which enter the amplitude only at NLO level.
TCS and DVCS amplitudes are identical (up to a complex conjugation) at lowest order in $\alpha_S$ but differ at next to leading order, in particular because of the quite different analytic structure of these reactions. Indeed the production of a timelike photon enables the production of intermediate states in some channels which were kinematically forbidden in the DVCS case. 
\subsection{LO estimates in ultraperipheral collisions}
We estimated \cite{PSW} the different contributions to the lepton pair cross section for 
ultraperipheral collisions at the LHC.  Since the
cross sections decrease rapidly with $Q'^2$, we are interested in the kinematics of moderate $Q'^2$, 
say a few GeV$^2$, and large energy, thus very small values of $\eta$. 
Note however that for a given proton energy the photon flux is higher at 
smaller photon energy.

\begin{figure*}
\includegraphics[width=0.5\textwidth]{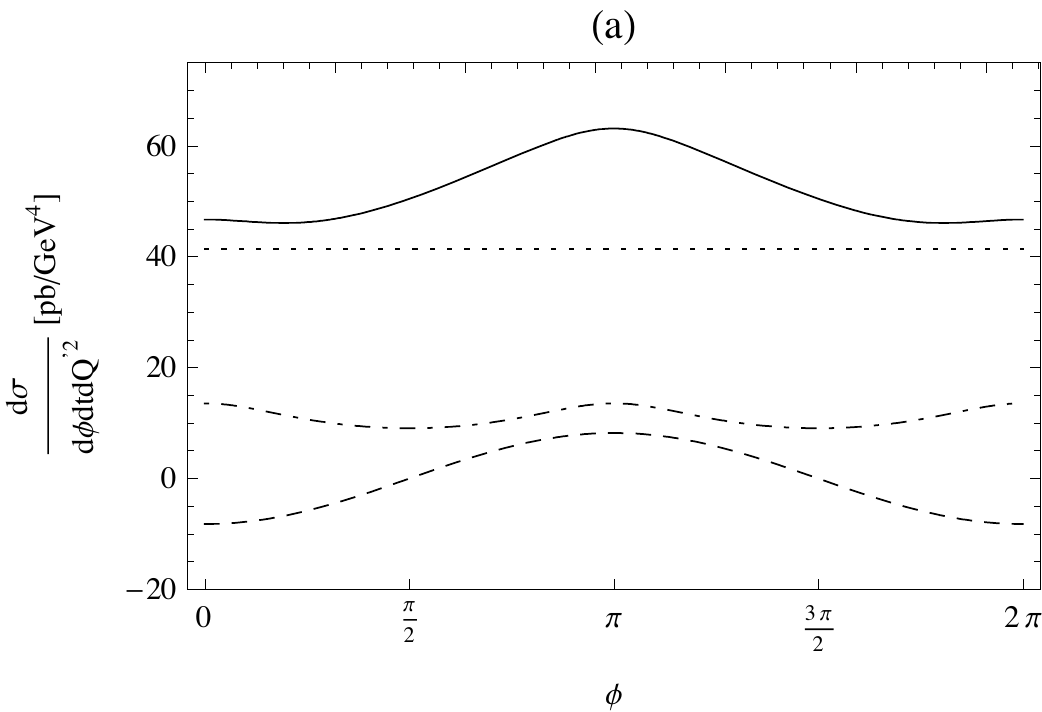}
\includegraphics[width=0.5\textwidth]{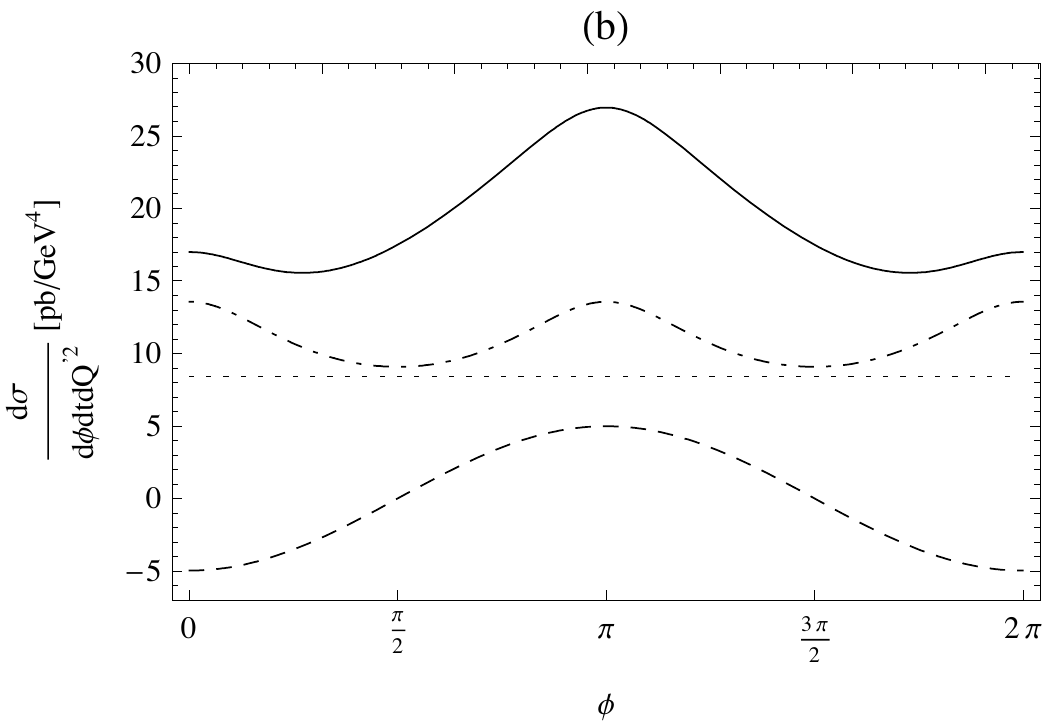} 
\caption{The differential cross sections (solid lines) for $t =-0.2 \gev^2$, ${Q'}^2 =5 \gev^2$  integrated 
over $\theta = [\pi/4,3\pi/4]$, as a function of $\varphi$, for $s=10^7 \gev^2$ (a), 
$s=10^5 \gev^2$(b)
 with $\mu_F^2 = 5 \gev^2$. We also display  the
Compton (dotted), Bethe-Heitler (dash-dotted) and Interference (dashed) contributions.}
\label{Interf}
\end{figure*}

In Fig. \ref{Interf} we show the interference contribution to the cross section in comparison to the Bethe Heitler and Compton processes, for various values of photon proton  energy 
squared $s = 10^7 \gev^2,10^5 \gev^2$. We observe that for 
larger energies the  
Compton process dominates, whereas for $s=10^5 \gev^2$ all contributions are comparable. 

 Our leading  order estimate shows that the factorization scale dependence of the amplitudes is quite high. This fact demands the understanding of higher order contributions with the hope that they will stabilize this scale dependence.

\subsection{NLO corrections}
Our calculations \cite{PSW2} of NLO corrections show  important differences between the coefficient functions describing the TCS case and  those describing DVCS. First, the $p^2 +i\varepsilon$ prescription for propagators turns into a  $\eta \to \eta+i\varepsilon$, rather than a  $\eta \to \eta-i\varepsilon$ as in the DVCS case. The second difference is the presence of minus signs under the logarithms, which  produce additional terms. Particularly $\log^2(-2-i\varepsilon)$ present in the TCS result may produce correction much bigger than the corresponding $\log^2(2)$ in the DVCS case. Another important difference between the DVCS and TCS amplitudes appear in their imaginary part, which  is present only in the DGLAP region for DVCS, while it is present in both DGLAP and ERBL regions for TCS. Defining the quark and gluon coefficient functions as
 \begin{eqnarray}
T^q = C_0^q+ C_1^q +\frac{1}{2} \log(\frac{ |Q^2|}{\mu_F^2}) C_{coll}^q ~~~~~; ~~~~~
T^g =  C_1^g +\frac{1}{2} \log(\frac{ |Q^2|}{\mu_F^2}) C_{coll}^g\;,
\nonumber
\label{eq:NLOTCSDVCS}
\end{eqnarray}
where $ C_0^q$ is the Born order coefficient function and $\mu_F$ is the factorization scale. $C^q_{coll} $ and $C^g_{coll} $ are directly related to the evolution equation kernels.

\begin{figure*}
\center
\includegraphics[width=0.4\textwidth]{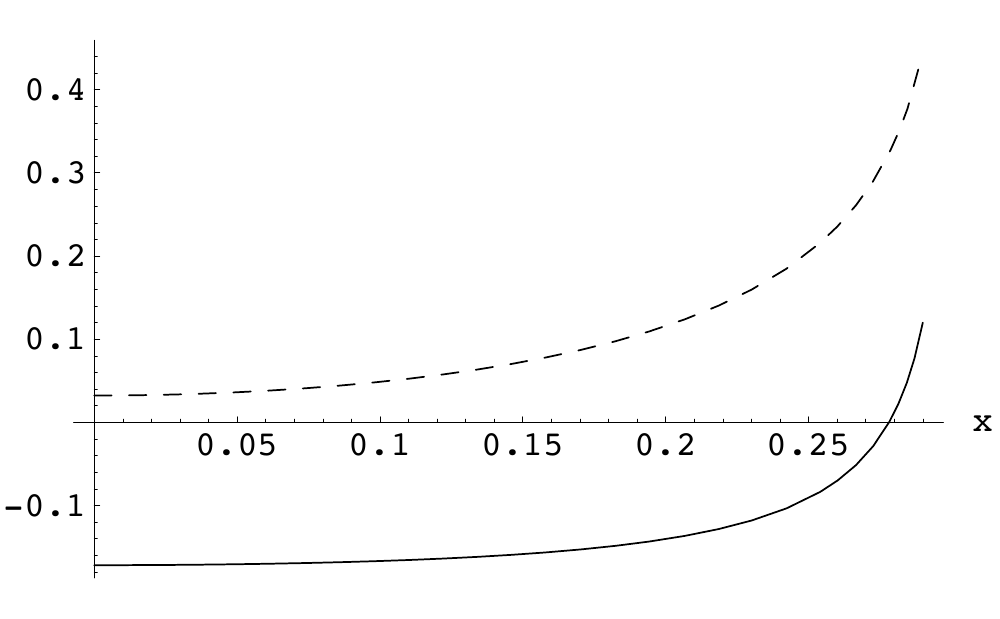} ~~~~~
\includegraphics[width=0.4\textwidth]{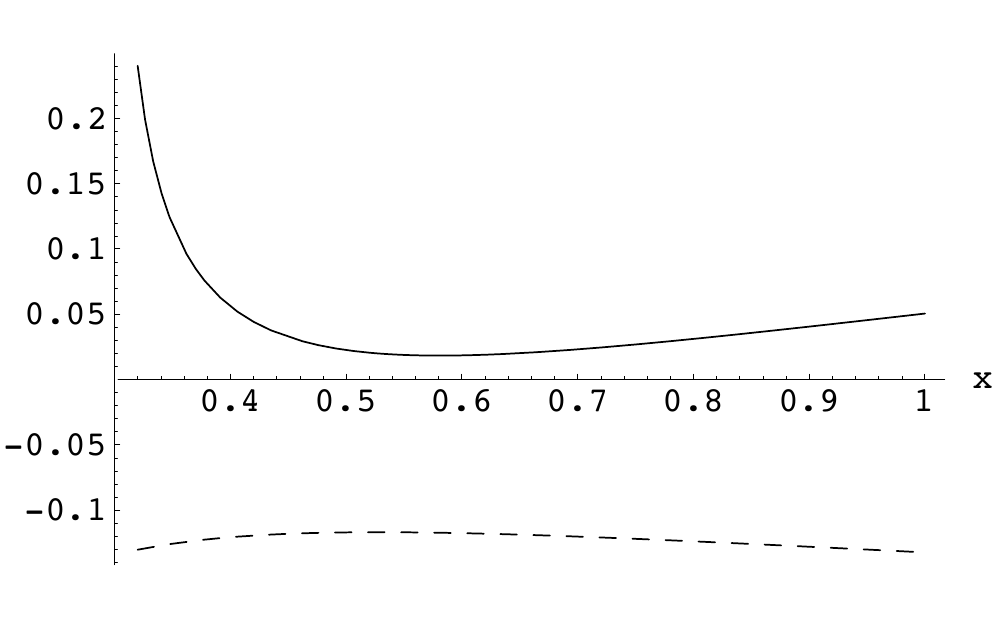} 
\includegraphics[width=0.4\textwidth]{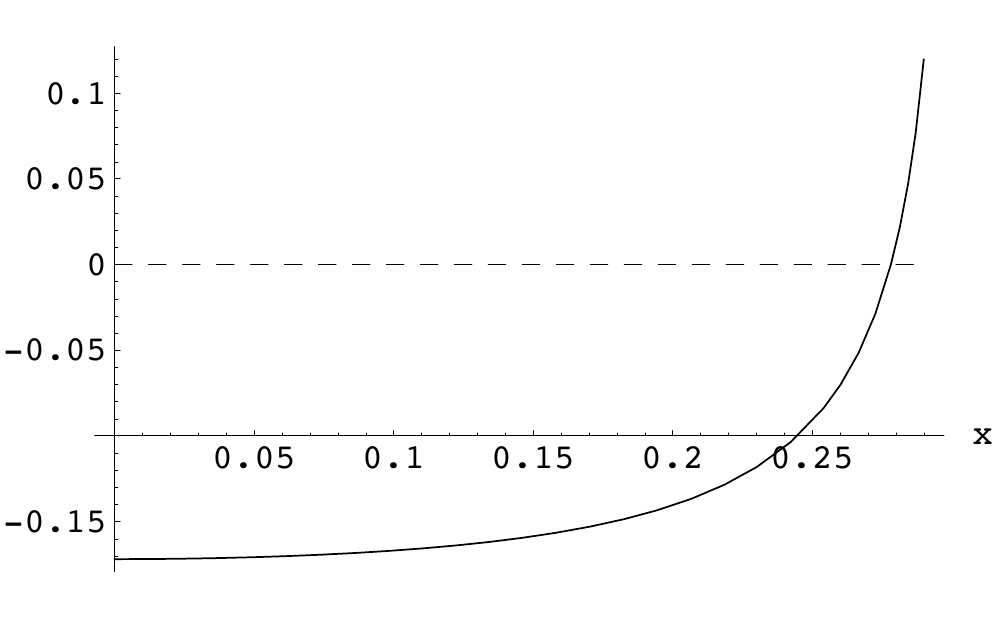} ~~~~
\includegraphics[width=0.4\textwidth]{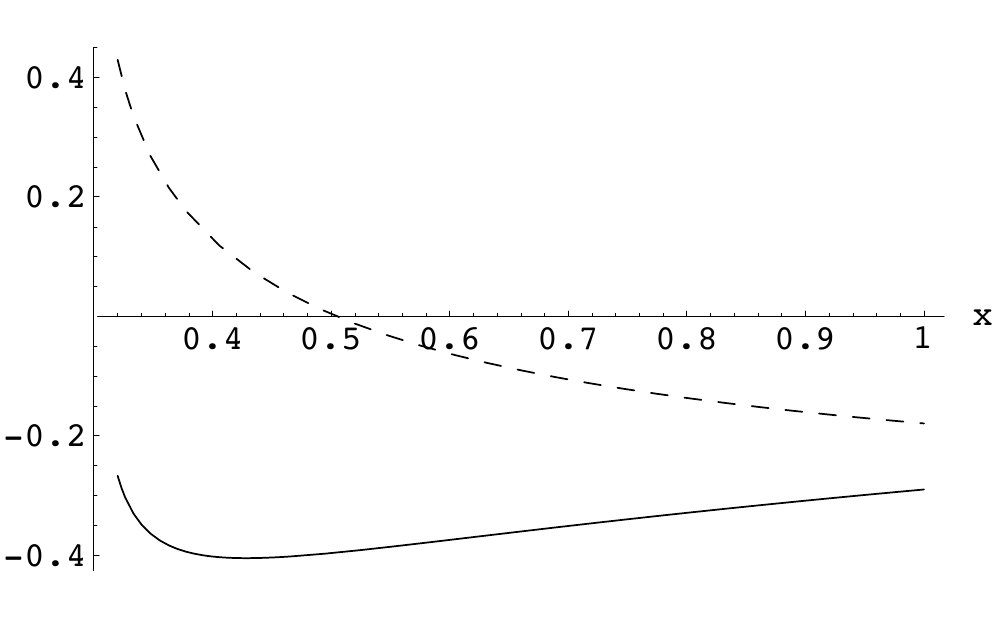} 
\caption{Real (solid line) and imaginary (dashed line) part of the ratio $R^q$ of the NLO quark coefficient function to the Born term in Timelike Compton Scattering (up) and Deeply Virtual Compton Scattering (down) as a function of $x$ in the ERBL (left) and DGLAP (right) region for $\eta = 0.3$, for $\mu_F^2 = |Q^2|$.}
\label{Fig:ratio}
\end{figure*}

To discuss the difference of the coefficient functions ${C_{1(TCS)}^q}^* - C_{1(DVCS)}^q  $ and present the magnitude of corrections we define the following ratio:
\begin{eqnarray}
R^q = \frac{C_{1}^q+\frac{1}{2}\log \left(\frac{|Q^2|}{\mu_F^2}\right) \cdot C^q_{coll}}{C^q_{0}}
\label{eq:r}
\end{eqnarray}
of the NLO quark correction to the coefficient function, to the Born level one. Let us restrict us to the factorization scale choice $\mu_F^2 = |Q^2|$.
On Fig. \ref{Fig:ratio} we  show  the real and imaginary parts of the ratio $R^q$ in timelike and spacelike Compton Scattering as a function of $x$ in the ERBL (left) and DGLAP (right) region for $\eta = 0.3$. We  fix $\alpha_s = 0.25$ and restrict the plots to the positive $x$ region, as the coefficient functions are antisymmetric in that variable. We see that in the TCS case, the imaginary part of the amplitude is  present in both the ERBL and DGLAP regions, contrarily to the DVCS case, where it exists  only in the DGLAP region. The magnitudes of these NLO coefficient functions are not small and neither is the  difference of the coefficient functions  ${C_{1(TCS)}^q}^* - C_{1(DVCS)}^q  $. The conclusion is that extracting the universal GPDs from both TCS and DVCS reactions requires much care.

\begin{figure*}
\hspace*{3cm} \includegraphics[width=0.4\textwidth]{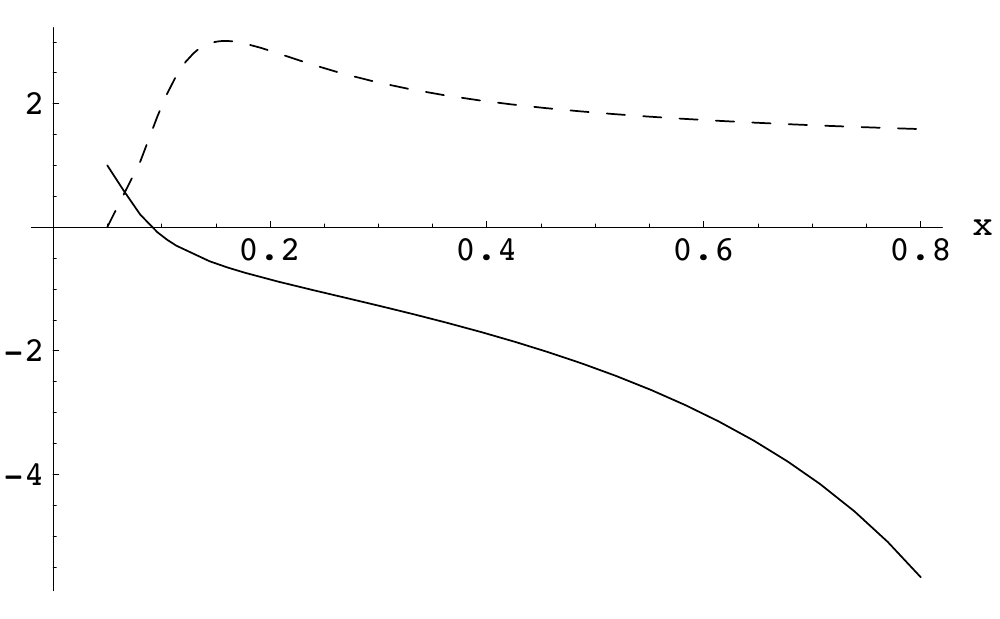}
\caption{Ratio of the real (solid line) and imaginary (dashed line) part of the NLO gluon coefficient function in TCS to the same quantity in DVCS as a function of $x$ in the DGLAP  region for $\eta = 0.05$ for $\mu_F^2 = |Q^2|$.
}
\label{Fig:ratiog}
\end{figure*}

Let us now briefly comment on  the gluon coefficient functions. 
The  real parts of the gluon contribution are equal for DVCS and TCS in the ERBL region. 
The differences between TCS and DVCS emerges in the ERBL region through the imaginary part of the coefficient function which is non zero only for the TCS case and is of the order of the real part. 
In Fig. \ref{Fig:ratiog} we plot the ratio 
$\frac{C^g_{1(TCS)} }{ C^g_{1(DVCS)}}$
of the NLO gluon correction to the hard scattering amplitude in TCS to the same quantity in the DVCS in the DGLAP  region for $\eta = 0.05$.

More phenomenological studies need now to be performed, by convoluting the coefficient functions to realistic quark and gluon GPDs and calculating the relevant observables in various kinematical domains. We are now  progressing on these points.

\section{On Transition Distribution Amplitudes}
Hadronic matrix elements of non-local light-cone operators are
the conventional non-perturbative objects which arise in the
description of hard exclusive electroproduction reactions
within the collinear factorization approach.
The factorization theorem for backward DVCS \cite{PS} and for hard exclusive backward meson electroproduction argued in
\cite{Frankfurt:1999fp,Frankfurt:2002kz}
lead to the introduction of baryon to meson transition distribution amplitudes (TDAs),
non diagonal matrix elements
of light-cone three quark operators
\begin{equation}
\widehat{O}^{\alpha \beta \gamma}_{\rho \tau \chi}(z_1,\,z_2,z_3)=
\varepsilon_{c_1 c_2 c_3}
\Psi^{c_1 \, \alpha}_\rho(z_1) \Psi^{c_2 \beta}_\tau (z_2) \Psi^{c_3 \, \gamma}_\chi (z_3)
\label{Def_operator}
\end{equation}
between baryon and meson states. In
(\ref{Def_operator})
$\alpha$, $\beta$, $\gamma$
stand for quark flavor indices;
$\rho$, $\tau$
and
$\chi$
denote the Dirac indices and
$c_{1,2,3}$
are indices of the color group.
If one adopts the light-cone
gauge
$A^+=0$,
 the gauge link is equal to unity and may be omitted
in the definition of the operator
(\ref{Def_operator}).
Baryon to meson TDAs extend
the concept of generalized parton distributions. They appear
as a building block in the colinear factorized description of amplitudes for a class of
hard exclusive reactions prominent examples being hard exclusive pion electroproduction off a nucleon
in the backward region and baryon-antibaryon annihilation into a pion and a lepton pair \cite{Pire:2005ax}.

\begin{figure}[htb!]
\includegraphics[width=0.3\textwidth,clip=true]{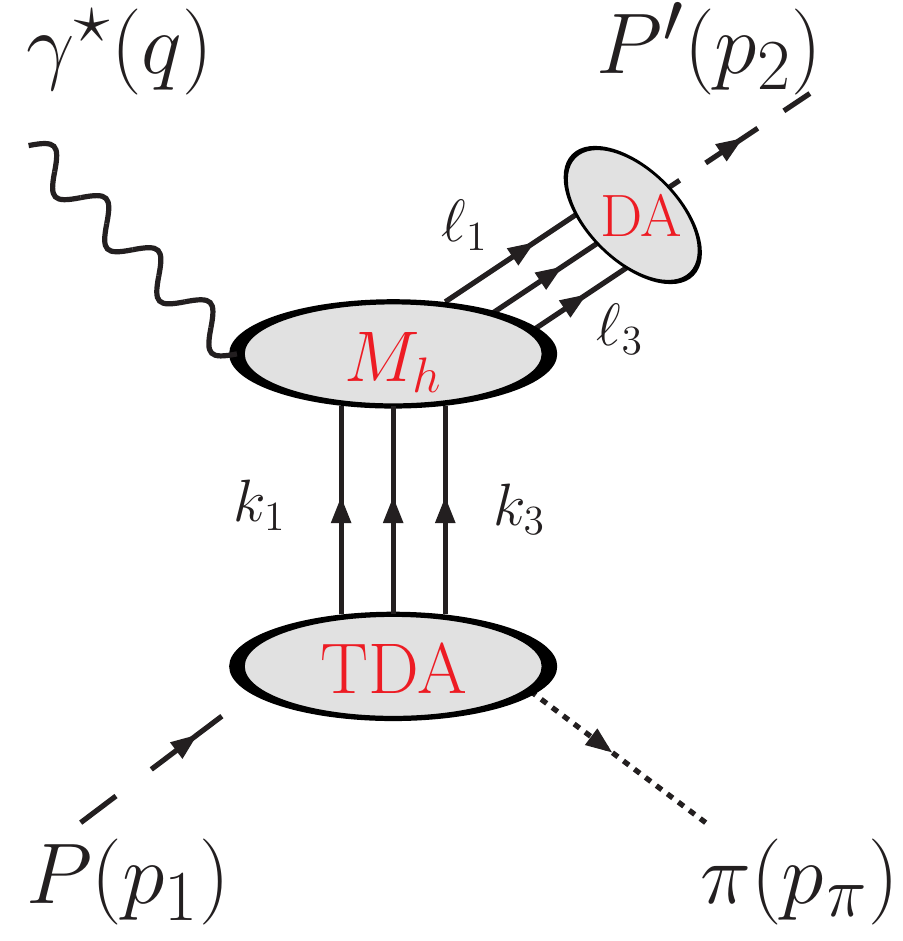}~~~
\includegraphics[width=0.3\textwidth,clip=true]{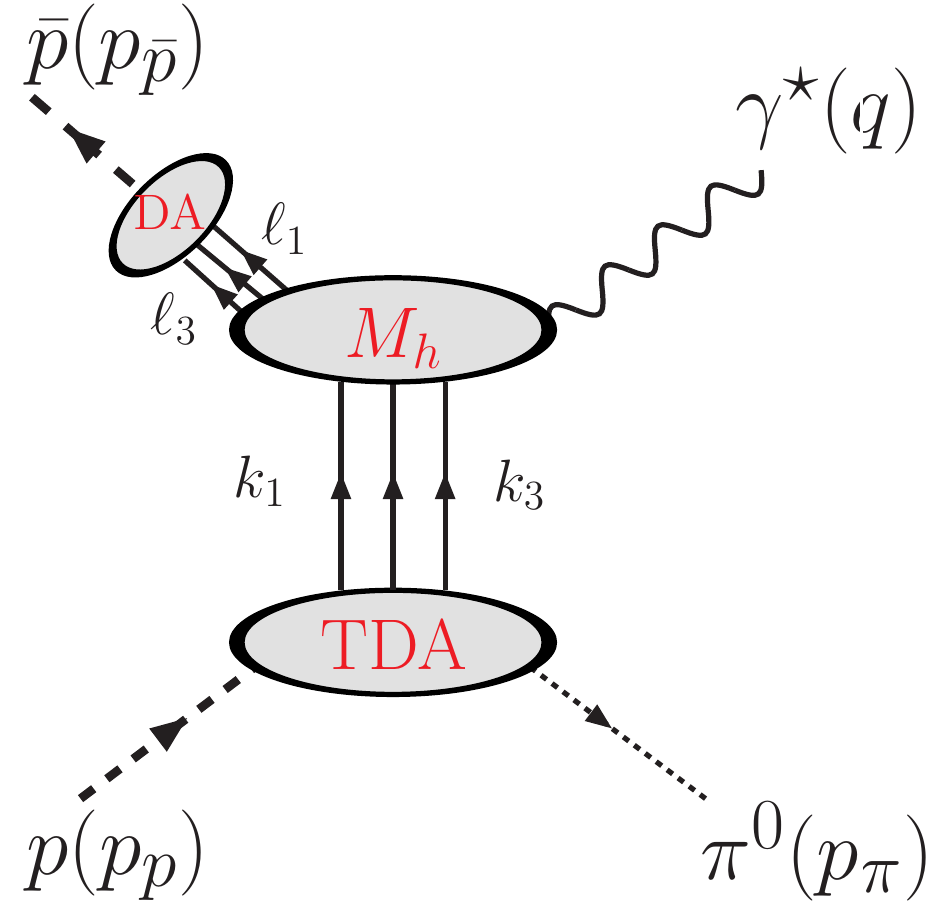}~~
\includegraphics[width=0.3\textwidth,clip=true]{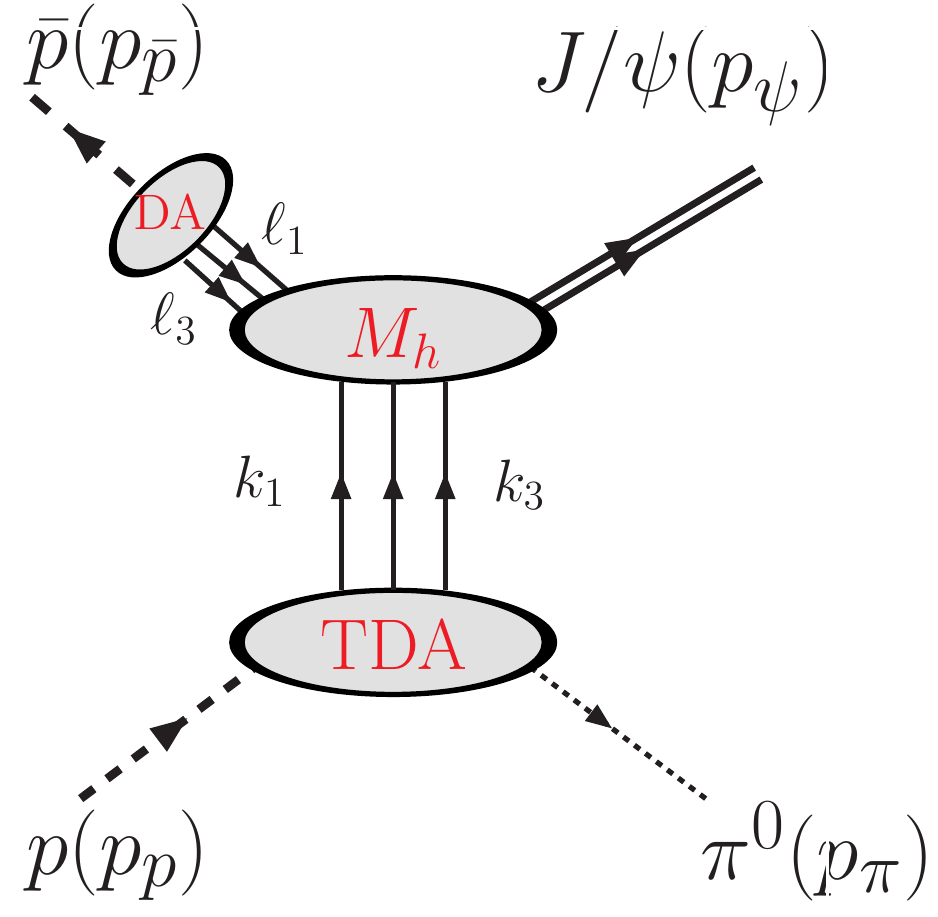}
\caption{Illustration of the factorisation for three exclusive reactions involving the TDAs.}
\label{fig:fact}
\end{figure}

$ \pi N$  TDAs appear in
the description of backward electroproduction of a pion on a nucleon target (see Fig.\ref{fig:fact} a). In terms of
angles, in the $\gamma^\star p$ center of momentum (CM) frame, the angle between
the $\gamma^\star$ and the pion, $\theta^\star_\pi$, is close to 180$^\circ$.
We then have $|u|\ll s$ and $t\simeq -(s+Q^2)$, in contrast to the fixed angle regime 
$u\simeq t\simeq -(s+Q^2)/2$ ($\theta^\star_\pi\simeq 90^\circ$) and the forward (GPD) one
$|t|\ll s$ and $u\simeq -(s+Q^2)$ ($\theta^\star_\pi\simeq 0^\circ$).

The TDAs appear also in similar electroproduction processes such 
as $e p \to  e\;(p,\Delta^+) \; (\eta,\rho^0)$,  $e p \to  e\;(n,\Delta) \; (\pi^+,\rho^+)$, $e p \to  e\;\Delta^{++}\; (\pi^-,\rho^-)$. Those processes have already been analysed, at
backward angles, at JLab in the resonance region, i.e. ~$\sqrt{s_{\gamma^\star p}}=W<1.8$ GeV, in order to study the baryonic 
transition form factors in the $\pi$ channel or in the $\eta$ 
channel. 
Data are being extracted in some channels above the 
resonance region. The number of events seems to be large enough to expect to get cross section measurements for 
$\Delta^2_T<1$ GeV$^2$, which is the region described in terms of TDAs. 
The HERMES analysis for forward electroproduction may  also be extended to larger values of  
$-t$.  It has to be noted though that present studies are limited to $Q^2$ of order a few GeV$^2$, 
which gives no guarantee to reach the TDA regime yet.
Higher-$Q^2$ data may be obtained at JLab-12 GeV and in muoproduction at Compass within the next few years.

Crossed reactions in proton-antiproton annihilation (e.g. with PANDA at GSI-FAIR~\cite{Lutz:2009ff}), 
with time-like photons (i.e.~di-leptons) and a pion (see Fig.\ref{fig:fact} b) also involve TDAs
both for small $t=(p_p-p_{\pi^0})^2$ and $u=(p_{\bar p}-p_{\pi^0})^2$. In the latter case, the TDAs
for a transition between the anti-proton and the pion are probed. One can also   study similar
reactions with other mesons than a pion, e.g.~$\bar p p \to  \gamma^\star\; (\eta,\rho^0) $, or on a different target
than proton $\bar p N \to \gamma^\star\pi$. Finally, one may also consider 
associated $J/\psi$ production with a pion  $\bar p p \to  J/\psi\; \pi^0 $ (see Fig.\ref{fig:fact} c) or another 
meson $\bar p p \to  J/\psi\; (\eta,\rho^0) $,
which involve the {\it same} TDAs as with an off-shell photon or in backward electroproduction.
They will serve as  very strong tests of the universality of TDAs in different processes.

The first application of baryonic TDAs was centered on backward electroproduction of a 
pion~\cite{LPS1}. In that case, the hard contribution which consists in the 
scattering of the hard photon with three quarks is known at leading order.
Extrapolating the limiting value of $\pi N$ TDAs obtained from the soft pion theorems, we obtained a first evaluation of the unpolarised cross section for 
backward electroproduction in the region of large-$\xi$.This estimate, which is unfortunately reliable only in a very
restricted kinematical domain (large-$\xi$), shows an interesting sensitivity to the underlying 
model of the proton DA. This study has been extended to hard exclusive production of a $\gamma^\star \pi^0$ pair in $\bar p p $ annihilation at GSI-FAIR~\cite{LPS2}.

$\pi N$ TDAs are non-perturbative objects governed by long distance dynamics.  
In accordance with the usual logic of the collinear factorization approach $\pi N$ TDAs 
have well established renormalization group properties.
The evolution properties of the three quark non-local operator on the light-cone
were extensively studied in the literature
for the case of matrix elements
between a baryon and the vacuum known as baryon distribution amplitudes (DAs).
Since TDAs involve the same operator its evolution determines the factorization scale dependence of TDAs
\cite{Pire:2005ax}. We derived \cite{Pire:2010if} a spectral representation for the $\pi N$ TDAs, and introduced the notion of quadruple distributions which enable to  generalize Radyushkin's factorized Ansatz from the GPD case to the TDA case.
 Lorentz invariance results in a polynomiality property of the 
Mellin moments of TDAs in the longitudinal momentum fractions.

The detailed account of isospin and permutation symmetries \cite{Pire:2011xv}
provides a unified description of all isotopic channels in terms of eight 
independent $\pi N$ TDAs. 
These general constraints derived  should be satisfied by 
any realistic model  of TDAs. 
The crossing relation between $\pi N$ TDAs and GDAs leads to 
a soft pion theorem for isospin-$\frac{1}{2}$ and isospin-$\frac{3}{2}$ $\pi N$ TDAs, which helps to derive normalization conditions for $\pi N$ TDAs. 
A simple resonance exchange model considering nucleon
and $\Delta(1232)$ exchanges in 
isospin-$\frac{1}{2}$ and isospin-$\frac{3}{2}$ channels respectively allows to approximate $\pi N$ TDAs in the ERBL region \cite{Pire:2011xv}. Nucleon exchange 
may be considered as a pure $D$-term contribution generating 
the highest power monomials of $\xi$ of the Mellin moments
in the longitudinal momentum fractions,  complementary to the spectral representation
for TDAs in terms of quadruple distributions.

Backward hard-exclusive reactions thus open  new windows in the understanding
of hadronic physics in the framework of the
collinear-factorization approach of QCD. To extract reliable precise information on the baryon to meson 
transition distribution amplitudes from an incomplete set
of observables such as cross sections and asymmetries, one needs
to develop  realistic models for the TDAs. Nonperturbative techniques such as lattice
simulations need to be adapted to this problem. Three approaches -- the soft pion theorem, the pion-cloud model  \cite{Barbara}
and the spectral representation -- are being explored and first 
cross-section evaluations in the whole kinematical domain covered
by the TDA factorisation should be available soon.

The 12 GeV JLab upgrade and the start-up of
GSI-FAIR hopefully will bring us the necessary experimental information
to test the model-independent predictions of the TDA factorisation
and then to check specific predictions from different TDA models potentially
connected to the more fundamental quantity of the hadronic realm.

\ack
We aknowledge useful discussions with Grzegorz Grzelak, Jean-Philippe Lansberg, Pawe{\l} Nadel-Turo\'nski and Samuel Wallon. This work was supported by the French-Polish Scientific Agreement Polonium and by the Consortium Physique des Deux Infinis (P2I).


\section*{References}
\begin{thebibliography}{9}


\bibitem{historyofDVCS}
M{\"u}ller  D {\em et al.} 1994
{\it Fortsch.\ Phys.} {\bf 42} 101

Ji X 1997
{\it Phys.\ Rev.\ Lett.} {\bf 78} 610

Radyushkin A V 1997
{\it Phys.\ Rev.} D {\bf 56}  5524 

Collins J C and Freund A 1999
{\it Phys.\ Rev.} D {\bf 59} 074009 

\bibitem{gpdrev}
Diehl M 2003
{\it
  Phys.\ Rept. }{\bf 388}  41
  
Belitsky A V and Radyushkin A V 2005
{\it
  Phys.\ Rept.} {\bf 418} 1 
 
  \bibitem{Burk}
Burkardt M 2000
{\it  Phys.\ Rev.} D {\bf 62} 071503; 
  
Ralston JP  and Pire B 2002
{\it  Phys.\ Rev.}  D {\bf 66} 111501 ;
  
  Diehl M 2002
{\it   Eur.\ Phys.\ J.} C  {\bf 25} 223 .
  
\bibitem{DVCSexp}
  A.~Airapetian {\it et al.}  [HERMES Collaboration]  2001
 {\it Phys.\ Rev.\ Lett.} {\bf 87} 182001 ;
 
Munoz Camacho C {\it et al.}  [Jefferson Lab Hall A Collaboration]  2006
 {\it Phys.\ Rev.\ Lett.} {\bf 97}  262002;

Chekanov S {\it et al.}  [ZEUS Collaboration]  2003
{\it  Phys.\ Lett.} B {\bf 573} 46;

Aktas A{\it et al.}  [H1 Collaboration] 2005
  {\it Eur.\ Phys.\ J.}  C {\bf 44} 1 ;

Stepanyan S {\it et al.}  [CLAS Collaboration]  2001
  {\it Phys.\ Rev.\ Lett.}  {\bf 87} 182002 .

\bibitem{fitting}
 Kumericki K, Mueller D and Passek-Kumericki K 2008
 {\it  Nucl.\ Phys.}  B  {\bf 794} 244 ;

  Guidal M and Moutarde H 2009
  {\it Eur.\ Phys.\ J.}  A {\bf 42} 71 ;
  
Moutarde H  2009
 {\it Phys.\ Rev.}  D  {\bf 79} 094021 .
 
\bibitem{APT}
 Anikin IV, Pire B and Teryaev OV  2000
  {\it Phys.\ Rev.} D {\bf 62} 071501 ;
  
Belitsky AV {\em et al.}  2001
  {\it Phys.\ Lett.} B  {\bf 510} 117 ;
  
Belitsky AV and Mueller D  2010
 {\it Phys.\ Rev.} D  {\bf 82} 074010 .





\bibitem{BDP}
Berger ER, Diehl M and Pire B 2002
  {\it Eur.\ Phys.\ J.} C {\bf 23} 675.

\bibitem{NadelTuronski:2009zz}
 Nadel-Turonski P {\em et al.}  2009
 { \it AIP Conf.\ Proc.} {\bf 1182} 843.

\bibitem{Albayrak}
Albayrak I {\em et al.}  
$e^+ e^-$ pair production with CLAS12 at 11 GeV,  CLAS letter of intent (2011).

\bibitem{PSW}
  Pire B, Szymanowski L and Wagner J  2009
  {\it Phys.\ Rev.} D {\bf 79} 014010 , {\it Nucl.\ Phys.\ Proc.\ Suppl.} {\bf 179-180} 232 
  and  {\it Acta Phys.\ Polon.\ Supp.} {\bf 2}, 373 .
 
 \bibitem{eic}
 Boer D {\it et al.} 2011
  Gluons and the quark sea at high energies: Distributions, polarization,
  tomography
   {\it Preprint} arXiv:1108.1713 [nucl-th].

 \bibitem{PSW2}
 Pire B, Szymanowski L and Wagner J  2011
  {\it Phys.\ Rev.} D {\bf 83}, 034009 .

  \bibitem{PS}
  Pire B and Szymanowski L 2005
  {\it Phys.\ Rev.} D {\bf 71} 111501 ;
  
 Lansberg JP,   Pire B and Szymanowski L 2006
  {\it Phys.\ Rev.} D {\bf 73} 074014 .
  
\bibitem{Frankfurt:1999fp}
  Frankfurt LL {\em et al.} 1999
  {\it Phys.\ Rev.} D {\bf 60} 014010.

\bibitem{Frankfurt:2002kz}
   Frankfurt LL {\em et al.}  2002
Novel hard semiexclusive processes and color singlet clusters in hadrons
   {\it Preprint} hep-ph/0211263.



\bibitem{Pire:2005ax}
  Pire B and Szymanowski L 2005
 {\it Phys.\ Lett.} B {\bf 622} 83 



\bibitem{LPS1}
  Lansberg JP, Pire B and Szymanowski L  2007
  {\it Phys.\ Rev.} D {\bf 75} 074004 
  [Erratum-ibid. D {\bf 77} 019902 ].
 
 
 \bibitem{LPS2}
  Lansberg JP, Pire B and Szymanowski L  2007
 { \it Phys.\ Rev.} D {\bf 76} 111502 .
  
  \bibitem{Lutz:2009ff}
Lutz MF  {\em et al.}  [The PANDA
                 Collaboration]  2009
 Physics Performance Report for PANDA: Strong Interaction Studies with Antiprotons
  {\it Preprint} arXiv:0903.3905 [hep-ex].


 
   \bibitem{Pire:2010if}
  Pire B, Semenov-Tian-Shansky K and Szymanowski L  2010
 { \it Phys.\ Rev.} D {\bf 82} 094030 ,
  
   Pire B, Semenov-Tian-Shansky K and Szymanowski L  2011
 { \it AIP Conf.\ Proc.} {\bf 1350} 237.
  
  \bibitem{Pire:2011xv}
  Pire B, Semenov-Tian-Shansky K and Szymanowski L  2011
  $\pi N$ transition distribution amplitudes: their symmetries and constraints from chiral dynamics
  {\it Preprint} arXiv:1106.1851 [hep-ph].

\bibitem{Barbara}
  Pasquini B, Pincetti M and Boffi S  2009
 { \it Phys.\ Rev.}  D {\bf 80} 014017 .
   

  
\end{thebibliography}
\end{document}